\documentclass[12pt]{article}
\usepackage[dvips]{graphicx}
\usepackage{cite}
\begin{document}
\title{Construction of the free energy landscape by the density functional
theory}
\author
{Takashi Yoshidome, Akira Yoshimori and Takashi Odagaki\\
 Department of Physics, Kyushu University, Fukuoka 812-8581, Japan}
\date{ }
\maketitle
\begin{abstract}
On the basis of the density functional theory, we give a clear definition of
the free energy landscape. To show the usefulness of the definition, we
construct the free energy landscape for rearrangement of atoms in an FCC
crystal of hard spheres. In this description, the cooperatively rearranging
region (CRR) is clealy related to the hard spheres involved in the saddle
between two adjacent basins.
A new concept of the simultaneously rearranging region (SRR) emerges naturally
as spheres defined by the difference between two adjacent basins. We show that
the SRR and the CRR can be determined explicitly from the free energy
landscape.
\end{abstract}

\newpage
\section{Introduction}
The thermodynamics and the dynamics of glass forming substances exhibit
anomalous behaviors near the glass transition temperature, $T_g$.
As the sample which was prepared by rapid cooling below $T_g$ is heated,
the specific heat exhibits an abrupt increment at
$T_g$\cite{hinetuexp_yamamuro}.
It is also known that supercooled liquids near $T_g$ show several
relaxation processes and that the relaxation time of the slowest process
tends to diverge at a temperature below $T_g$\cite{rev_glass}.
It is, therefore, desirable to construct a theory for the glass transition
which gives a unified explanation for these dynamic and thermodynamic
anomalies.

The mode coupling theory (MCT)\cite{MCT} for the glass transition
describes the liquid dynamics based on the generalized Langevin equation
and predicts an ergodic-to-nonergodic transition at a critical temperature,
$T_c$ due to the nonlinear coupling between various modes. However,
it is now believed that $T_c$ is much higher than $T_g$.
In addition, MCT cannot treat the thermodynamic singularities.
Mezard and Parisi estimated thermodynamic properties of glassy state
in quasi-equilibrium with the replica method \cite{Mezard}.
In this approach, the Kauzmann temperature, $T_K$, becomes an ideal glass
transition temperature. However, $T_K$ is much lower than $T_g$ and
the replica method cannot treat the dynamics either.
Both theories thus can not give a unified explanation for the thermodynamic
and dynamic singularities of supercooled liquids near $T_g$.
  
We have proposed the free energy landscape picture by which 
a unified understanding is given for the singularities
near $T_g$\cite{odagaki04}.
The free energy landscape is defined in the configurational space
as a function of fictitious localized atomic positions and
each point in the free energy landscape is determined by coarse-graining of
the microscopic motion in the configurational space.
The free energy landscape is expected to have many basins and the saddle
points at low temperatures. The separation of the dynamics can be
understood by the separation of the dynamics among basins and the dynamics
within a basin. 
The thermodynamic and dynamic singularities near $T_g$ can also be explained
by the free energy landscape picture. The trapping diffusion model\cite{TDM}
and its generalization to dynamics in a landscape\cite{odagaki04} provide a unified
understanding of the dynamical singularities near the glass transition.
This model showed that the dynamical transition occurs at the temperature
where the mean waiting time for escaping from a basin diverges.
Tao et al.\cite{hinetu_tao} proposed a model based on the free energy
landscape picture to explain the specific heat anomaly at the glass transition.
They showed that the anomaly of the specific heat occurs when the mean waiting
time exceeds the observation time at low temperatures. 

It is now believed that the glass transition can be phenomenologically well
explained by the free energy landscape picture. It is thus an important
question how one can explicitly define and construct the landscape
on the basis of the microscopic Hamiltonian. Because of this lack of
an explicit theory, the analysis of the dynamics in the landscape has
been detered so far.

The aim of the present paper is to define explicitly the free energy landscape
using the microscopic Hamiltonian. To this end, we exploit the density
functional theory (DFT)\cite{HM,Oxtoby,Singh,Yoshimori}.

One of the key concepts to understand the glass transition singularity
is the cooperatively rearranging region (CRR) proposed by Adam and
Gibbs\cite{AG}. 
Adam and Gibbs succeeded in explaining the non-Arrenius behavior in the
temperature dependence of the viscosity for fragile glass forming liquids.
However there is no clear common understanding of the physical meaning for
the CRR. From the present definition of the free energy landscape,
the clear meaning of the CRR emerges. In addition to the CRR,
the present construction of the landscape leads to a new concept of
rearranging region, namely simultaneously rearranging region (SRR)
which is defined by the difference between the atomic
configuration in two adjacent basins.

This paper is organized as follows. We give a definition of the free energy
landscape by the DFT in Sec 2. To confirm the usefulness of this definition,
we construct the free energy landscape for rearrangement of atoms in an FCC
crystal of hard spheres in Sec 3. A relation between the free energy landscape
and the CRR is given in Sec 4. We also discuss the SRR in Sec. 4.
Results are summarized in Sec 5.
\section{Definition of the free energy landscape}
We define the free energy by the partition function which is obtained by
the partial sum of the fast motion in the configurational space.
The free energy becomes a function of the fictitious atomic positions
and we call this free energy as the landscape.
We exploit the density functional theory (DFT)\cite{HM,Oxtoby,Singh,Yoshimori}
to calculate the free energy. In the DFT the grand potential,
$\Omega[\rho(\textbf{r})]$, is expressed as a functional of the density
field, $\rho(\textbf{r})$, and the equilibrium density field,
$\rho_{eq}(\textbf{r})$, is determined as a minimum of the grand potential.
If the density field is given as a function of a particle configuration,
$\{ \textbf{R}_i \} \equiv \{\textbf{R}_1,\textbf{R}_2,\cdots,\textbf{R}_N \}$
where $\textbf{R}_i$ is the position of the \textit{i}-th particle,
the density functional $\Omega[\rho(\textbf{r})]$ becomes a function of
$\{ \textbf{R}_i \}$ which can be considerd as the free energy landscape
defined above.
We employ a sum of Gaussians as the density field:
\begin{eqnarray}
\rho(\textbf{r})=\Bigl(\frac{\alpha}{\pi} \Bigr)^{\frac{3}{2}} \sum_i 
\exp[-\alpha(\textbf{r}-\textbf{R}_i)^2].
\label{eq:gauss_hennbunn}
\end{eqnarray}
Here $\alpha$ and $\textbf{R}_i$ are the degree of the spread of the density
distribution and the position of the \textit{i}-th particle, respectively.
This density field has been used for the investigations such as the
liquid-solid transition\cite{Oxtoby,Singh,Barrat, Mohanty} and the glass
transition\cite{Wolynes1,Baus,Lowen,Das,Kim}.
The density field (\ref{eq:gauss_hennbunn}) means that the distribution of
the vibrational motion around $\{ \textbf{R}_i \}$ is approximated by Gaussian
functions. The vibrational motion within
$| \textbf{r}-\textbf{R}_i| \le 1/\sqrt{\alpha}$ is coarse-grained by $\alpha$.
By using eq. (\ref{eq:gauss_hennbunn}), $\Omega[\rho(\textbf{r})]$ becomes a
function of $\alpha$ and $\{\textbf{R}_i \}$:
\begin{eqnarray}
\Omega[\rho(\textbf{r})] \equiv \Omega(\alpha, \{ \textbf{R}_i \}).
\label{eq:def_fel}
\end{eqnarray}

We use the grand potential (\ref{eq:def_fel}) as a definition of the free
energy landscape. Our definition of the free energy landscape is different
from the definition by Dasgupta and Valls \cite{Dasgupta_FEL}. They defined
the free energy landscape as a functional of the density field,
$\rho(\textbf{r})$.
\section{Calculation of the free energy landscape}
\subsection{Model}
We use the approximation proposed by Ramakrishnan and Yussoff to calculate
$\Omega[\rho(\textbf{r})]$\cite{RY}. The result of the approximation is as
follows:
\begin{eqnarray}
\beta \Delta \Omega[\rho(\textbf{r})]&\equiv&\beta \Omega[\rho(\textbf{r})]-
\beta \Omega[\rho_l] \nonumber \\
&=& \int_{V} d \textbf{r} \rho(\textbf{r}) \log\Bigl[\frac{\rho(\textbf{r})}
{\rho_l} \Bigr]-\int_V d \textbf{r} (\rho (\textbf{r})-\rho_l) \nonumber \\
& &-\frac{1}{2} \int_V d \textbf{r}_1 \int_V d \textbf{r}_2 c(|\textbf{r}_1-
\textbf{r}_2|)(\rho (\textbf{r}_1)-\rho_l)(\rho (\textbf{r}_2)-\rho_l) .
\label{eq:RY_basic}
\end{eqnarray}
Here $\beta=(k_BT)^{-1}$ where $k_B$ is the Boltzmann constant and $T$ is the
temperature, and $\rho_l$ denotes the density of the uniform liquid.
The coefficient $c(|\textbf{r}|)$ is the direct correlation function of the
uniform liquid at $\rho_l$.

We substitute eq. (\ref{eq:gauss_hennbunn}) into eq. (\ref{eq:RY_basic}):
\begin{eqnarray}
\beta \Delta \Omega(\alpha,\{\textbf{R}_i\}) &=& F_{id}(\alpha,
\{\textbf{R}_i\})+F_0-\frac{N_s}{2} I_0(\alpha)-\frac{1}{2} \sum_{i}
\sum_{j \neq i} I(\alpha,\textbf{R}_{ij}) 
\label{eq:RY_koeki_alpha_R}
\end{eqnarray}
where
\begin{eqnarray}
& & F_{id}(\alpha,\{\textbf{R}_i\})= \Bigl(\frac{\alpha}{\pi}
\Bigr)^{\frac{3}{2}} \sum_i \int_V d \textbf{r} \exp[-\alpha(\textbf{r}-
\textbf{R}_i)^2] \log \left \{\frac{1}{\rho_l} \Bigl(\frac{\alpha}{\pi}
\Bigr)^{\frac{3}{2}} \sum_j \exp[-\alpha(\textbf{r}-\textbf{R}_j)^2]
\right \} \nonumber \\ \\
& & F_0=N_l \biggl( \rho_s-\frac{\rho_l}{2} \biggr) \int_V d \textbf{r} c(r) \\
& & I_0(\alpha)=\Bigl(\frac{\alpha}{\pi} \Bigr)^3 \int_V d \textbf{r}_1 
\int_V d \textbf{r}_2 c(|\textbf{r}_1-\textbf{r}_2|) 
\exp[-\alpha \textbf{r}_1^2] \exp[-\alpha \textbf{r}_2^2] \\
\label{eq:I0}
& & I(\alpha,\textbf{R})=\Bigl(\frac{\alpha}{\pi} \Bigr)^3 \int_V d 
\textbf{r}_1 \int_V d \textbf{r}_2 c(|\textbf{r}_1-\textbf{r}_2|) 
\exp[-\alpha \textbf{r}_1^2] \exp[-\alpha(\textbf{r}_2+\textbf{R})^2] .
\label{eq:IR}
\end{eqnarray}
The average density $\rho_s$ is defined by 
\begin{eqnarray}
\rho_s=\frac{1}{V} \int_V d \textbf{r} \rho(\textbf{r}) ,
\end{eqnarray}
and $N_s \equiv \rho_s V$, $N_l \equiv \rho_l V$.

In the present paper, we treat a system of the hard spheres.
We use Percus-Yevick approximation for the direct correlation function of the
hard sphere \cite{HM}:
\begin{eqnarray}
c(r)&=&
\left \{
\begin{array}{lr}
B_0+B_1r+B_2r^3 &(r < \sigma) \\
0 & (r > \sigma). \\
\end{array}
\right.
\label{eq:PY}
\end{eqnarray}
where $\sigma$ is the diameter of the hard sphere.
The coefficients $B_0$, $B_1$, and $B_2$ in eq. (\ref{eq:PY}) are given by
\begin{eqnarray}
B_0&=&-\frac{(1+2 \eta)^2}{(1-\eta)^4} \\
B_1&=&6 \eta \frac{(1+\eta/2)^2}{(1-\eta)^4} \\
B_2&=&-\frac{\eta}{2} \frac{(1+2 \eta)^2}{(1-\eta)^4},
\end{eqnarray}
and $\eta=\pi \rho_l/6$. By using eq. (\ref{eq:PY}), $F_0$, $I_0(\alpha)$,
and $I(\alpha,\textbf{R})$ can be calculated
analytically\cite{Mohanty,keisannhouhou}. 

We use $\alpha \sigma^2=478$ correponding to the liquid-solid transition
point \cite{Barrat}. For large $\alpha$, $F_{id}(\alpha,\{\textbf{R}_i\})$
reduces to
\begin{eqnarray}
F_{id}(\alpha,\{\textbf{R}_i\}) & \simeq & N_s
\Bigl(\frac{\alpha}{\pi})^{\frac{3}{2}} \int_V d\textbf{r} 
\exp[-\alpha \textbf{r}^2] \log 
\Bigl[\frac{\exp[-\alpha \textbf{r}^2]}{\rho_l} 
\Bigl(\frac{\alpha}{\pi}\Bigr)^{\frac{3}{2}} \Bigr] \nonumber \\
&=&N_s \Bigl[\frac{3}{2}\log
\Bigl(\frac{\alpha}{\pi}\Bigr)-\log(\rho_l)-\frac{5}{2} \Bigr]+\rho_l.
\label{eq:L_Fid}
\end{eqnarray}
Though Jones and Mohanty proposed that $I_0(\alpha)$ and $I(\alpha,\textbf{R})$
also reduce to more simple form \cite{Mohanty} for large $\alpha$, we do not
use their expression, since it can be used only for large $\alpha$, and is
not applicable for the entire region of $\alpha$. We use the analytical form
which can be used for all $\alpha$ \cite{keisannhouhou}.
\subsection{Forced relaxation in an FCC lattice}
In order to show the usefulness and the validity of the present definition
of the free energy landscape, we calculate the free energy landscape for a
forced structural change in an FCC crystal of hard spheres.
Since the grand potential (\ref{eq:def_fel}) has
too many parameters, $\alpha$ and $\{ \textbf{R}_i \}$, we calculate the free
energy landscape for a process in which we force to move a specific group of
particles and to relax other particles to minimize
$\Omega(\alpha,\{\textbf{R}_i\}) $ in the $\{ \textbf{R}_i \}$ space.
We calculated the free energy landscape when we force to rotate two adjacent
particles with their distance fixed (Figure 1).
The axis of rotation is the center of the two particles and two particles are
rotated around an axis in $<100>$ direction.
The arrangement of other particles is changed to minimize
$\Omega(\alpha,\{\textbf{R}_i\}) $ in the $\{ \textbf{R}_i \}$ space with
the positions of the two particles fixed.
We exploit the steepest decent method to find the minimum.

We imposed periodic boundary conditions in all directions.
The number of the particles in the system is 108.
We set $\rho_s \sigma^3=1.0$ and $\rho_l \sigma^3=0.967$ which correspond to
the metastable state to the liquid state\cite{Barrat}.
Figure 2 shows the free energy landscape as a function of rotation angle
$\theta$. The free energy landscape has a saddle point at $\theta=\pi/2$ and
two minima at $\theta=0$ and $\theta=\pi$. Two minima at $\theta=0$ and
$\theta=\pi$ correspond to the FCC configuration (Figures 3 (a) $\sim$ (e)).
The height of the saddle point is about $57k_BT$.
\section{CRR and SRR in the free energy landscape}
\subsection{Definition of CRR and SRR}
We define the simultaneously rearranging region (SRR) as the difference in the
configuration at the minimum between the two adjacent basins which are connected directly via a saddle point. Apparently, the number of the particles in SRR is
2 in the system studied in Sec 3.

The CRR corresponds to the particles which are moved at the saddle point
from the configuration at the minimum of the original basin.
Our definition of the CRR is thought to be closely related to the idea of the
CRR by Adam and Gibbs\cite{AG}, since the particles which are moved at the
saddle point are thought to give the configurational entropy for the
rearrangement proposed by Adam and Gibbs\cite{AG}.
\subsection{Calculation of the CRR}
Since we use $\alpha \sigma^2=478$, the degree of the coarse-graining is about
$0.05 \sigma$. Hence we assumed that the particle in the CRR for the system
calculated in Sec 3 is the particle displacing more than $0.05 \sigma$ at
$\theta=0.5 \pi$ from the position at $\theta=0$. 

We found that the number of the particles in the CRR is $46$.
The particles which are shown by the bold circles at $\theta=\pi/2$ in
Figure 3 are those which satisfy the criterion.
\section{Summary}
In the present paper, we have given a clear definition of the free energy
landscape using the density functional theory.
As a test of the definition of the free energy landscape, we have calculated
the free energy landscape for rearrangement of atoms in an FCC crystal of
hard spheres.

We have clarified the concept of the cooperatively rearranging region (CRR) by
giving a relation between the free energy landscape and the CRR. We also have
introduced a new region, simultaneously rearranging region (SRR) which is
different from the CRR. We have calculated the number of the particles in the
SRR and the CRR from the calculated free energy landscape.

The present frame work can be easily generalized for glass-forming substances.
Hence we can calculate the number of the particles in the CRR and the height
of the saddle point near the glass transition from the free energy landscape.
Details will be studied in a forthcoming paper.

\section*{Acknowlegdement}
This work was supported in part by the Grant-in-Aid for Scientific Research
from the Ministry of Education, Culture, Sports, Science and Technology.

\newpage
\begin{center}
{\Large Figure Captions}
\end{center}
\begin{description}
\item[Fig 1] 
 A schematic representation of the model under consideration. 
Two particles are rotated with their distance fixed. The arrangement of other
 particles are changed to minimize $\Omega(\alpha,\{\textbf{R}_i\}) $ in
the $\{ \textbf{R}_i \}$ space.
\item[Fig 2]
The free energy landscape when we force to rotate the two particles.
The horizontal axis is rotation angle $\theta$. The vertical axis is the
difference between $\Omega(\alpha, \{ \textbf{R}_i \})$ of a certain rotation
angle and $\Omega(\alpha, \{ \textbf{R}_i \})$ of FCC. The number of the
particles in the system is 108. We set $\rho_s \sigma^3=1.0$, 
$\rho_l \sigma^3=0.967$, and $\alpha \sigma^2=478$. Although we write the
horizontal axis by one dimension, the actual free energy landscape has
 $108 \times 3$ dimensions.
\item[Fig 3]
The configurations at the plane which two particles rotate. Each figure
corresponds to (a) $\theta=0$, (b) $\theta=0.3\pi$, (c)
 $\theta=0.5\pi$, (d) $\theta=0.7\pi$, and (e) $\theta=\pi$.
Two particles with the bold circle are the particles rotating.
The particles with the bold circle at $\theta=0.5\pi$ are the particles in CRR.
\end{description}
%
\begin{figure}[h]
\vspace{1cm}
\begin{center}
\fbox{Fig 1}
\end{center}
\vspace{1cm}
\includegraphics[width=15cm,clip]{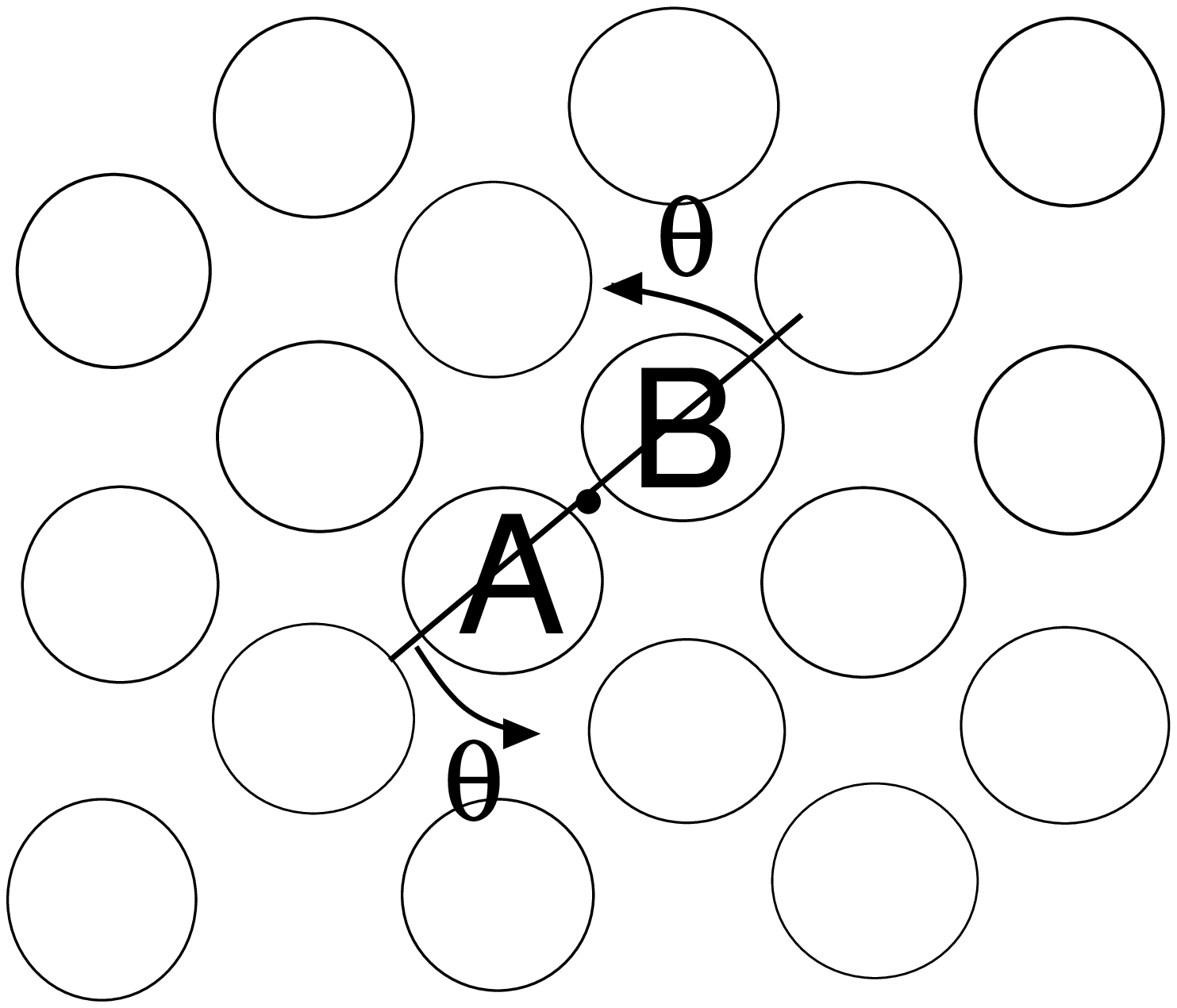}
\end{figure}
%
\begin{figure}[h]
\vspace{1cm}
\begin{center}
\fbox{Fig 2}
\end{center}
\vspace{1cm}
\includegraphics[width=15cm,clip]{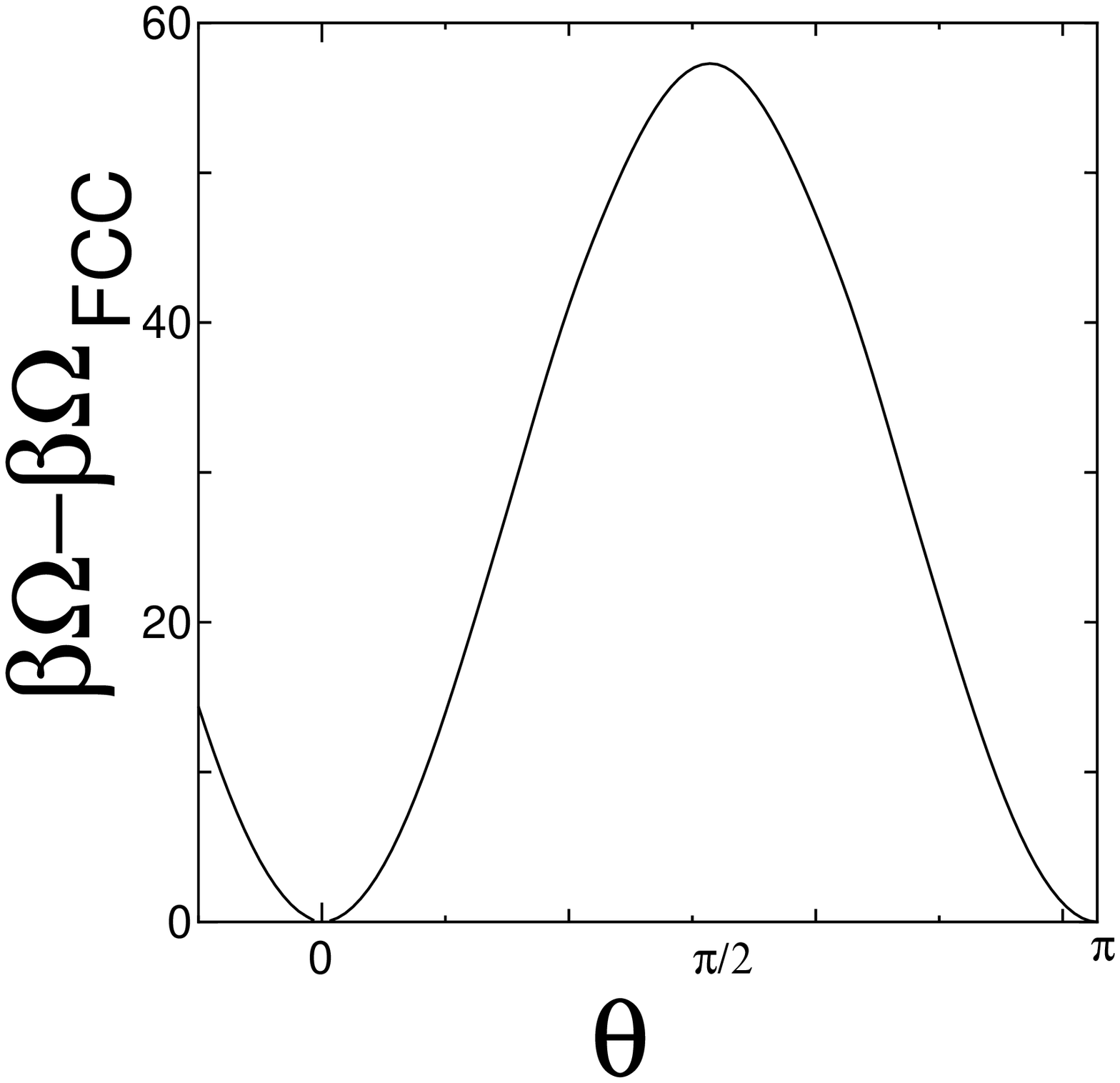}
\end{figure}
%
\begin{figure}[htbp]
\begin{center}
\fbox{Fig 3}
\end{center}
\begin{center}
\vspace{1cm}
\begin{minipage}{6.0cm}
\includegraphics[width=5cm,clip]{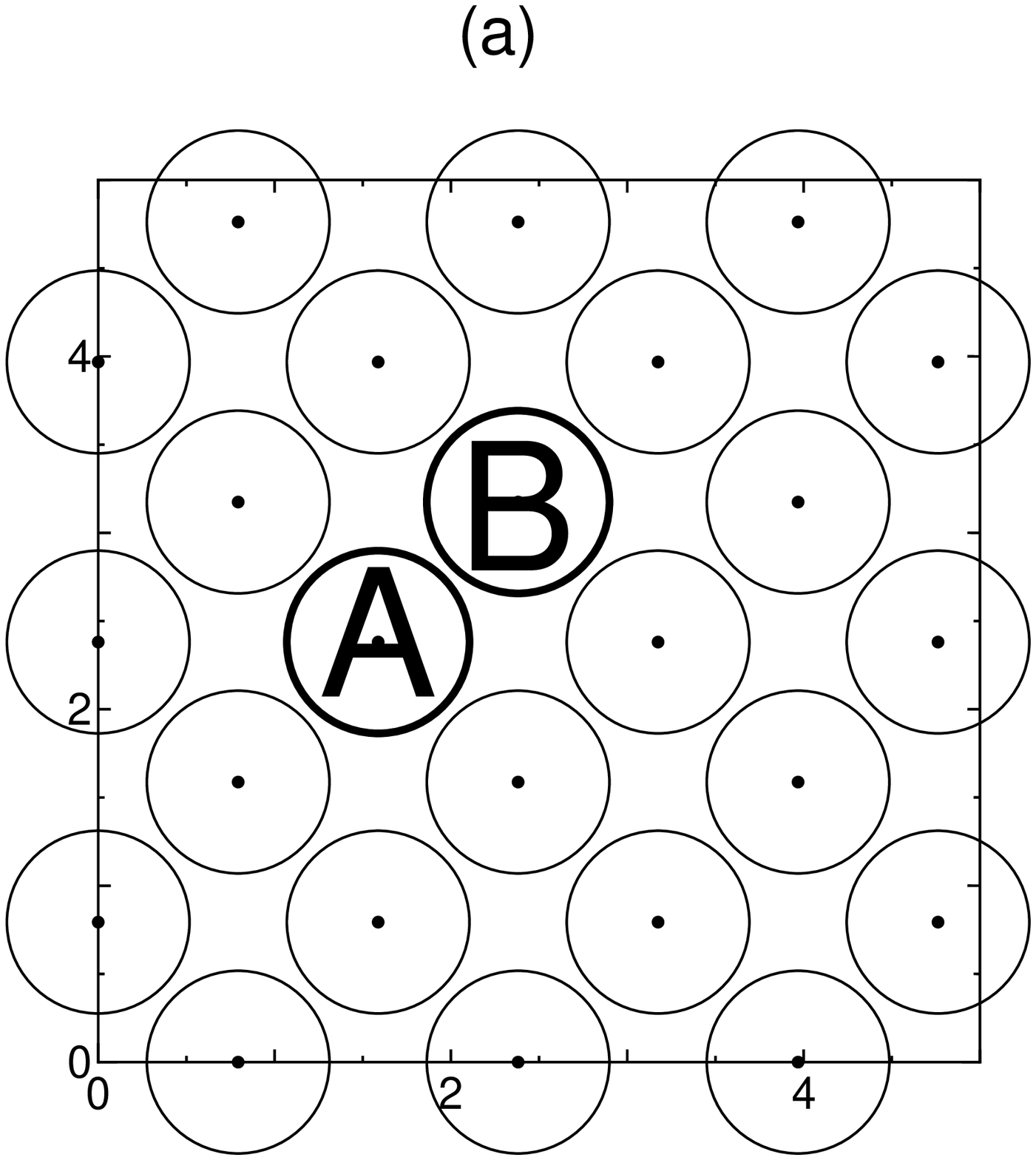}
\vspace{1cm}
\end{minipage} 
\hspace{1cm}
\begin{minipage}{6.0cm}
\includegraphics[width=5cm,clip]{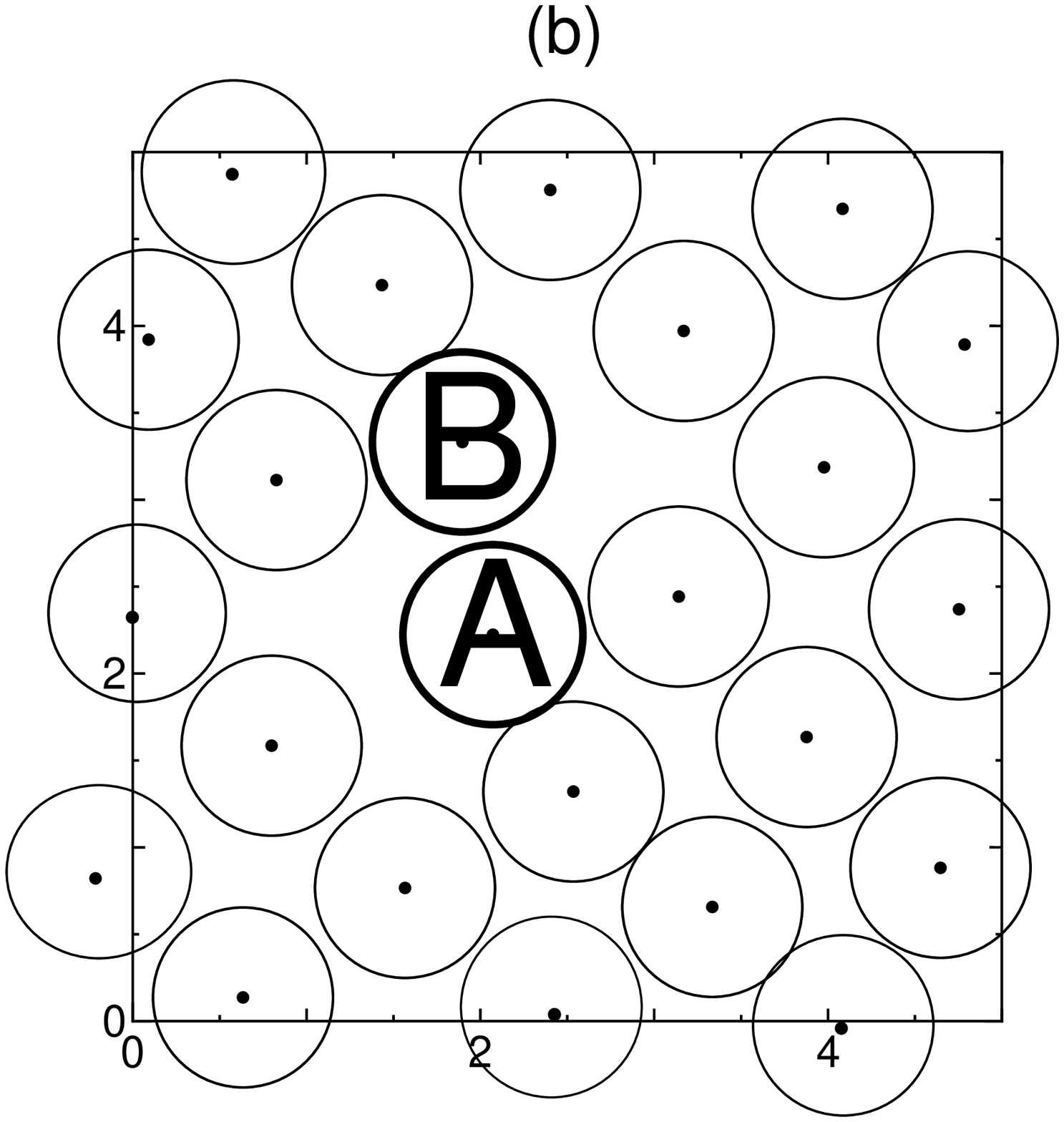}
\vspace{1cm}
\end{minipage}
\begin{minipage}{6.0cm}
\includegraphics[width=5cm,clip]{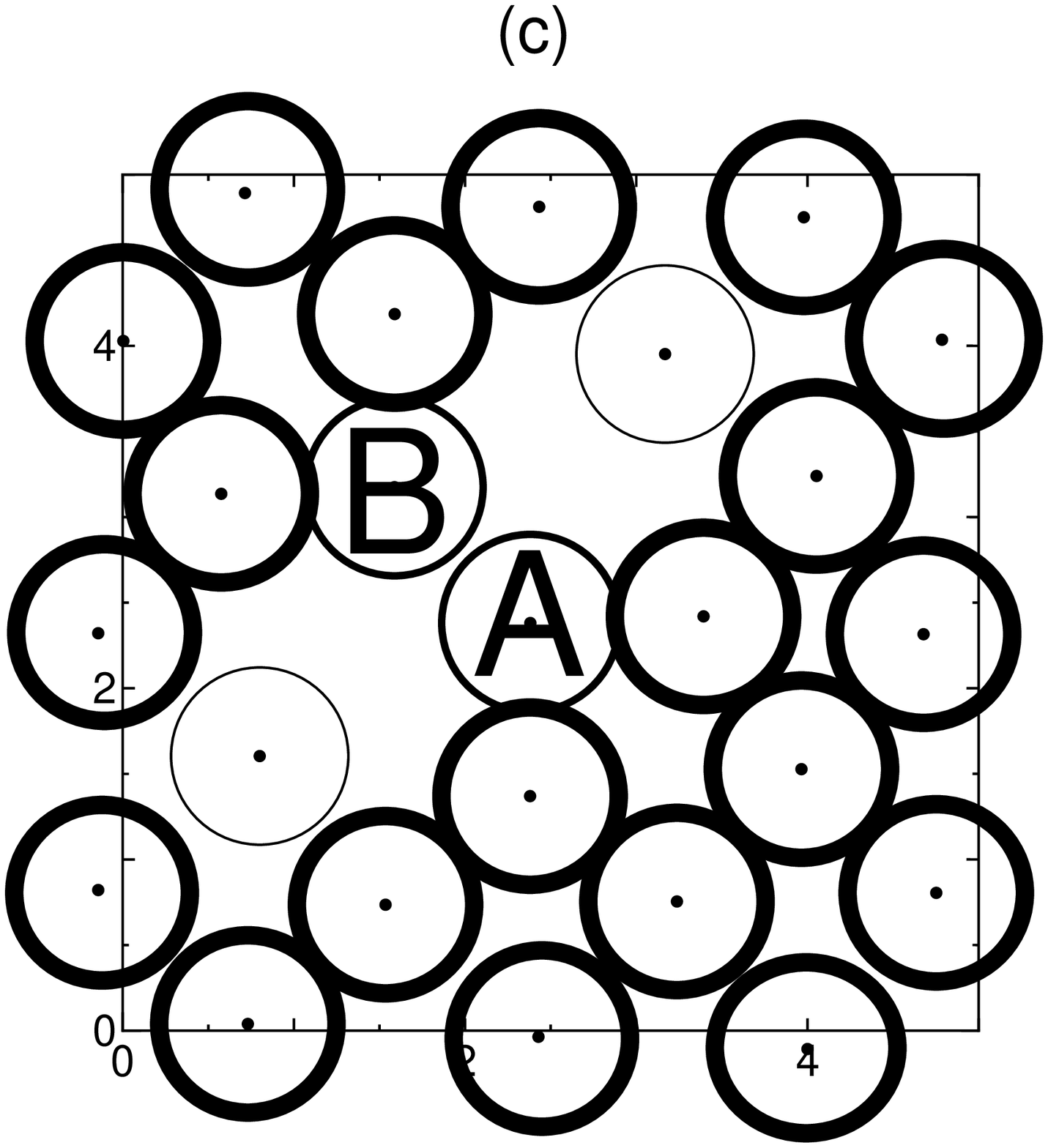}
\vspace{1cm}
\end{minipage} 
\hspace{1cm}
\begin{minipage}{6.0cm}
\includegraphics[width=5cm,clip]{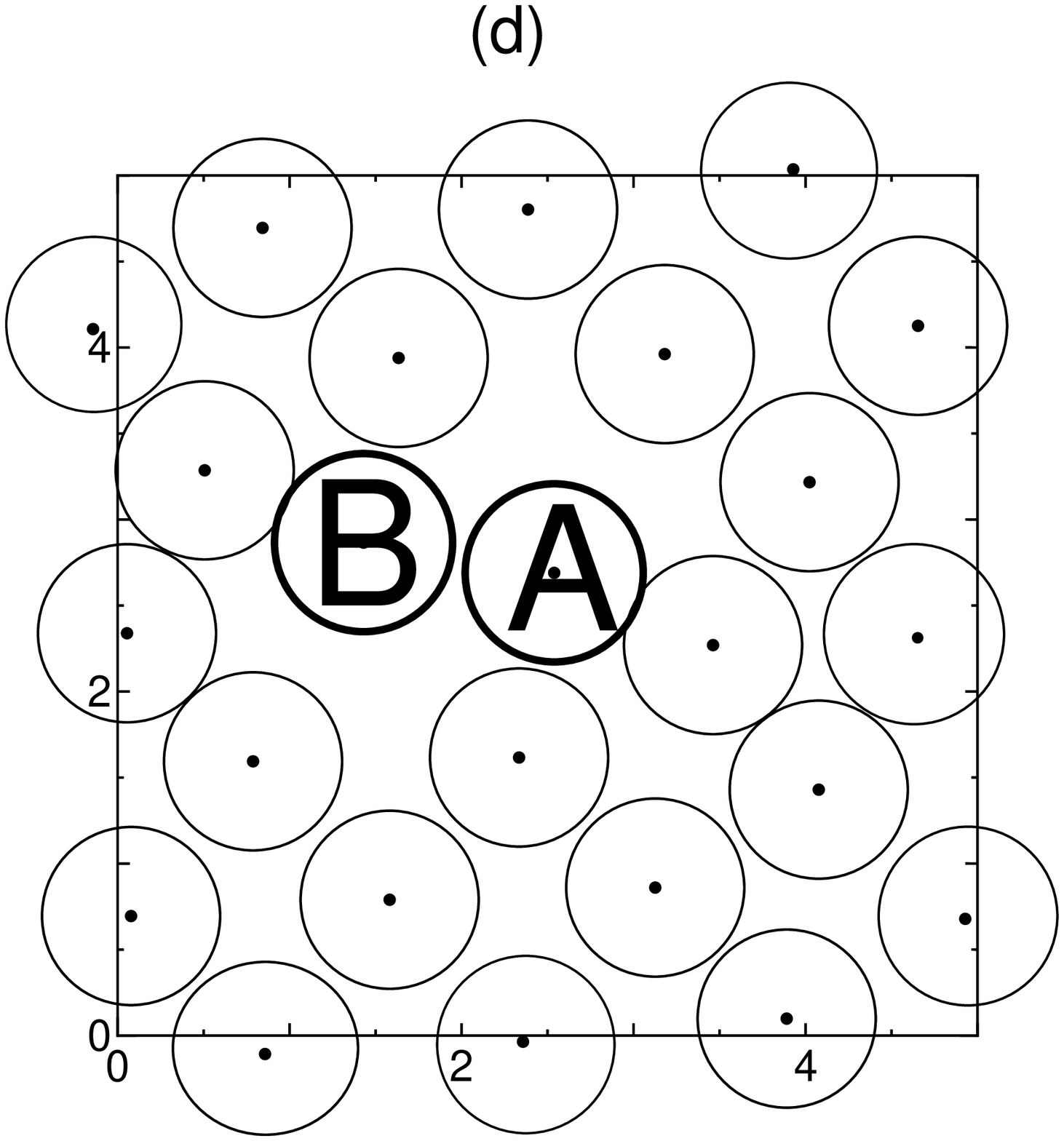}
\vspace{1cm}
\end{minipage}
\includegraphics[width=5cm,clip]{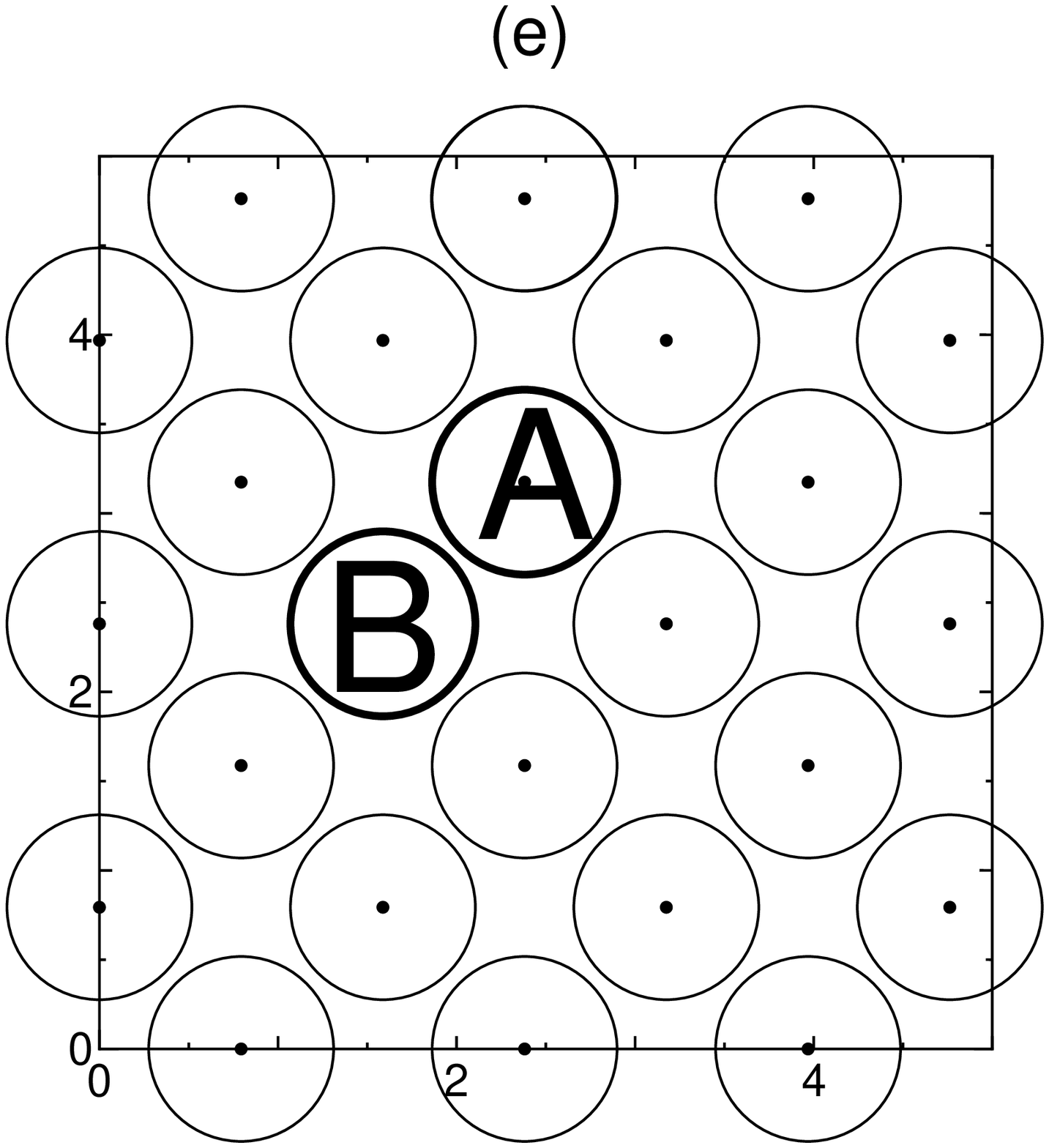}
\end{center}
\end{figure} 
\end{document}